\documentclass[a4paper]{article}

\usepackage{ifpdf}
\usepackage{cite}
\usepackage[pdftex]{graphicx}
\graphicspath{{./}}
\DeclareGraphicsExtensions{.pdf,.jpeg,.png}

\usepackage{xspace}

\usepackage[cmex10]{amsmath}
\usepackage{amssymb}
\usepackage{array}
\usepackage{mdwmath}
\usepackage{mdwtab}
\usepackage{eqparbox}
\usepackage[caption=false,font=footnotesize]{subfig}
\usepackage{fixltx2e}
\usepackage{stfloats}
\usepackage{url}
\usepackage{tikz}

\usepackage{RR}

\def\signed #1{{\leavevmode\unskip\nobreak\hfil\penalty50\hskip2em
  \hbox{}\nobreak\hfil#1%
  \parfillskip=0pt \finalhyphendemerits=0 \endgraf}}

\newsavebox\mybox
\newenvironment{aquote}[1]
  {\savebox\mybox{\emph{#1}}\begin{quote}}
  {\signed{\usebox\mybox}\end{quote}}

\newcommand{\neighbors}{\mathcal{N}}
\newcommand{\cohesion}{\mathcal{C}}
\newcommand{\tin}{\triangle_i}
\newcommand{\tout}{\triangle_o}

\newcommand\FEnbUsers{2557\xspace}
\newcommand\FEnbMaleUsers{1797\xspace}
\newcommand\FEnbFemaleUsers{698\xspace}
\newcommand\FEnbUnknownUsers{62\xspace}
\newcommand\FEnbTotalUsers{2635\xspace}
\newcommand\FEnbFailedUsers{78\xspace}
\newcommand\FEnbEgomunities{67750\xspace}
\newcommand\FEavAge{29.31\xspace}
\newcommand\FEstdAge{8.99\xspace}
\newcommand\FEavAgeM{29.76\xspace}
\newcommand\FEstdAgeM{8.80\xspace}
\newcommand\FEavAgeF{28.03\xspace}
\newcommand\FEstdAgeF{9.24\xspace}
\newcommand\FEnbFriendsTenPC{74\xspace}
\newcommand\FEnbFriendsFiftyPC{237\xspace}
\newcommand\FEnbFriendsNinetyPC{581\xspace}
\newcommand\FEnbRated{51161\xspace}
\newcommand\FEpercRatingOne{25.1\%\xspace}
\newcommand\FEpercRatingTwo{21.8\%\xspace}
\newcommand\FEpercRatingThree{22.5\%\xspace}
\newcommand\FEpercRatingFour{30.7\%\xspace}
\newcommand\FEpercRatedAtLeastNinetyPC{78\%\xspace}
\newcommand\FEunratedCohesionAv{0.108\xspace}
\newcommand\FEunratedCohesionStd{0.107\xspace}

\newcommand\FEcorCohRatSpearman{0.90\xspace}
\newcommand\FEcorCohRatSpearmanP{9.1\times 10^{-37}\xspace}

\newcommand\FEcorlogCohlogRatPearson{0.97\xspace}
\newcommand\FEcorlogCohlogRatPearsonP{2\times 10^{-61}\xspace}

\RRetitle{ Triangles to Capture Social Cohesion }
\RRtitle{ Des Triangles pour Capturer la Coh\'esion Sociale}
\RRdomaine{3}
\RRtheme{R\'eseaux et t\'el\'ecommunications}
\RRauthor{
  Adrien Friggeri
  \and
  Guillaume Chelius
  \and
  Eric Fleury
}
\URRhoneAlpes
\RRequipe{DNET}
\RRdate{8 July 2011}
\RRNo{}

\RRmotcle {r\'eseaux sociaux, r\'eseaux complexes, graphes r\'eels, d\'etection de communaut\'es, communaut\'es recouvrantes, data mining, mod\'elisation}

\RRabstract{
Although community detection has drawn tremendous amount of attention across
the sciences in the past decades, no formal consensus has been reached on the
very nature of what qualifies a community as such. In this article we take an
orthogonal approach by introducing a novel point of view to the problem of
overlapping communities. Instead of quantifying the quality of a set of
communities, we choose to focus on the intrinsic community-ness of one given
set of nodes. To do so, we propose a general metric on graphs, the cohesion,
based on counting triangles and inspired by well established sociological
considerations. The model has been validated through a large-scale online
experiment called Fellows in which users were able to compute their social
groups on Facebook and rate the quality of the obtained groups. By observing
those ratings in relation to the cohesion we assess that the cohesion is a
strong indicator of users subjective perception of the community-ness of a set
of people.
}

\RRresume{ 
Bien que la probl\'ematique de d\'etection de communaut\'es dans les r\'eseaux sociaux
ait attir\'e une attention grandissante \`a travers les sciences ces derni\`eres
ann\'ees, aucun consensus formel n'a \'et\'e atteint sur la nature de ce qui d\'efinit
une communaut\'e. Nous introduisons ici un point de vue novateur au probl\`eme de
communaut\'es recouvrantes. Au lieu de quantifier la qualit\'e d'un ensemble de
communaut\'es, nous nous concentrons sur l'aspect intrins\`equement communautaire
d'un ensemble donn\'e de nœuds. Pour ce faire, nous proposons une m\'etrique
g\'en\'erique sur les graphes, la coh\'esion, se fondant sur la notion de triangles
et inspir\'ee par des r\'esultats \'etablis en sociologie. Ce mod\`ele a \'et\'e valid\'e \`a
travers Fellows, une exp\'erience \`a large \'echelle sur Facebook dans laquelle les
utilisateurs avaient la possibilit\'e de calculer de mani\`ere automatique leurs
groupes d'amis puis de noter la qualit\'e de ceux ci. En observant ces notes et
la coh\'esion des groupes obtenus, nous concluons que la coh\'esion est une bonne
\'evaluation de la perception subjective de l'aspect communautaire d'un ensemble
de n\oe uds par un utilisateur.
}

\begin{document}
\RRNo{7686}
\makeRR

\section*{Introduction}
\begin{aquote}{The Social Science Encyclopedia, \emph{Adam Kuper}}
The term community relates to a wide range of phenomena and has been used as
an omnibus word loaded with diverse associations.
\end{aquote}

Although community detection has drawn tremendous amount of attention across
the sciences in the past decades, no formal consensus has been reached on the
very nature of what qualifies a community as such. In 1955, George Hillery,
Jr. analyzed 94 different sociological definitions of the term
\emph{community} \cite{Hillery:fqmsdk} both from a quantitative and
qualitative standpoint only to conclude that their only common defining
feature was that they all dealt with people. Despite this fact, there were
other traits on which the majority of definitions agreed, and he stated that
\emph{``of the 94 definitions, 69 are in accord that social interaction, area,
and a common tie or ties are commonly found in community life''}. Fast-forward
half a century, through the emergence of network science in the last two
decades, the \emph{communities community} has expanded to encompass scientists
coming from backgrounds as diverse as, among others, computer science,
theoretical physics or biology who brought along their own ideas and baggage
on what should be called a community.

In this context, where social networks are modeled as graphs of individuals
linked when they share a social connection in real life, all authors concur on
the intuitive notion that a community is a relatively tightly interconnected
group of nodes which somehow features less links to the rest of the network.
Unfortunately, this agreement does not extend to the specific formal meanings
of \emph{tightly interconnected} and \emph{less links}. The important aspect
to consider, however, is that the defining concept of community in network
science resides in topological features of the network. In real life, however,
one rarely describes a group of people as \emph{``this set of 10 people of
density 0.8, featuring on average 2 outbound link per individual''},
understandably preferring clearer -- and yet less formal -- labels such as
\emph{`family'}, \emph{`people at work'} or \emph{`the poker group'}.

The whole idea behind community detection in social networks is due to the
observation that there is a correlation between the topology of the network
and some kind of labels which relate to social interactions\footnote{For
obvious reasons, this assertion does not hold for arbitrarily defined groups,
consider for example the set of people of even height or any other group
sharing a randomly distributed feature: chances that this group present a
distinctive topological structure which separates it from the rest of the
network are pretty slim.}, and that therefore it should be possible to infer
the socio-semantic structure of the network by observing some of its
topological traits.

Social networks are a peculiar beast in the sense that they only exist as
descriptions of a fragment of what one would call \emph{The Social Network},
an unmeasurable, exhaustive and dynamic multigraph of all social interactions
at mankind scale. For example, Zachary's famous karate club
dataset\cite{zachary77} is nothing more than Zachary's description of a subset
of all social interactions, limited in terms of people (members of a karate
club in a US university), nature (friendship) and time (at some point in time
in the 1970s).

It is therefore important to keep in mind that any structural properties of
communities are constrained by the nature of the network. The emergence of
online social networks such as Facebook and Twitter in the last years and the
availability of high computational power has led to a unique situation where
there are not only rich datasets to study but also the ability to do so. But
the richness of these networks lead them not only shine by their size but also
their intricate complexity, as they encompass social links which may vary both
in nature and in intensity. For example, people add close friends as well as
professional acquaintances on Facebook, treating both categories as equals --
in Facebook's terms, all social links are friendships -- effectively
flattening a complex multi-graph into a very slightly less complex graph.

In that case, what meaning should one give to \emph{less links}? It is
obvious, for example, that excluding an employee's boss from their ``family''
group should not be detrimental to the group's community-ness, whereas
excluding their mother should. And yet in both cases the topological
implications are the same: a edge in the network links someone inside the
group to someone outside. Thus, given that all links are not equal in the
network, the considered topological features should go beyond the simple
notion of edges in order to discriminate those type of cases.

In this article, we introduce in Section \ref{sec:Cohesion} the
\emph{cohesion}, a new graph metric, inspired by well established sociological
results, which rates the intrinsic community-ness of a set of nodes of a
social network, independently from the existence of other communities. We then
describe in Section \ref{sec:Fellows} the experimental setup of Fellows, a
large scale online experiment on Facebook which we launched to prove the
validity of the cohesion. Finally in Section \ref{sec:Experimental Validation}
we exhibit the high correlation between the cohesion of social groups and the
subjective perception of those groups by users.


\section{Cohesion} 
\label{sec:Cohesion}

Before introducing the cohesion, let us reflect on the way community detection
has blossomed in the past few years. In 2004, at the junction of graph
partitioning in graph theory and hierarchical clustering in sociology, Newman
and Girvan proposed an algorithm to partition a network into several
communities. In order to assess the quality of the partitions which were
produced by their algorithm, they introduced the modularity
\cite{Newman:2004te}, a quantity which measures ``\emph{the fraction of the
edges in the network that connect vertices of the same type (i.e.,
within-community edges) minus the expected value of the same quantity in a
network with the same community divisions but random connections between the
vertices.}''

In the following years, the modularity attracted attention, with several
heuristics being proposed to attempt to find maximal partitions
modularity-wise -- see for example the Louvain method \cite{Blondel:2008vn}.
During the same time, other have exhibited several shortcomings of the
modularity itself: that it has a resolution limit and therefore that
modularity optimization techniques cannot detect small communities in large
networks, that some random networks are modular.

Going further, when partitioning a network, each node is affected to a unique
community, which has the rather unfortunate side effect of tearing families
apart: an individual cannot be at the same time part of \emph{their family}
and \emph{their company}.

In order to overcome these limitations, it is natural to shift to a context of
overlapping communities, in which the one-node-to-one-community constraint
disappears. This however has an incidence on modularity. ``\emph{If vertices
may belong to more clusters,}'' says Fortunato in his 2010 review, ``\emph{it
is not obvious how to find a proper generalization of modularity. In fact,
there is no unique recipe.}'' Naturally, other techniques such as clique the
percolation method \cite{Palla:2005ub} which do not rely on modularity were
introduced -- clique percolation goes even further as the method does not
evaluate the quality of communities.

Behind the beautiful simplicity of the modularity actually lie two subtly
different measures. First, the modularity encompasses the individual and
intrinsic quality of each community's \emph{content} by comparing them to a
null model. Second, but no less important, it implicitly judges the quality
of the \emph{division} in communities. While this makes sense in the context
of a partition because both those aspects are linked -- one cannot change the
content of a community without affecting other communities -- there is no
equivalent notion in an overlapping context.

\subsection{A Word on Judging Divisions} 
\label{sub:a_word_on_judging_divisions}
Judging the quality of the division largely depends on the data one wishes to
study. While it is obvious that two completely disjoint communities $S_1\cap
S_2 = \emptyset$ form a good division of the network $(S_1\cup S_2, E)$ and
that two completely overlapping communities $S_1=S_2=S$ form a really bad
division of the network $(S,E)$, the intermediate overlapping cases are less
trivial.

On the one hand, in some occurrences, there is a case for allowing small
\emph{fuzzy} overlaps in order to model an vertex-based interface between
groups instead of purely edges. On the other hand, there also are extreme
cases where communities should be allowed to overlap at a great extent --
consider for example college classes -- or even be allowed to be fully
embedded one in another (\emph{e.g.} a computer science lab might be a small
community inside a bigger university community).

For those reasons, we assess that their is no swiss army knife of division
rating: the tools used to rate the division in communities itself should be
carefully crafted to fit the data analysis.


\subsection{Rating the Content of Communities} 
\label{sub:definition}

It is however possible to rate the quality of one given community embedded in
a network, independently from the rest of the network. The idea is to give a
score to a specific set of nodes describing wether the underlying topology is
\emph{community like}. In order to encompass the vastness of the definitions
of what a community is, we propose to build such a function, called
\emph{cohesion}, upon the three following assumptions:
\begin{enumerate}
  \item the quality of a given community does not depend on the collateral
  existence of other communities;
  \item nor it is affected by remote nodes of the network;
  \item a community is a ``dense'' set of nodes in which information flows
  more easily than towards the rest of the network.
\end{enumerate}
The first point is a direct consequence of the previously exhibited dichotomy
between content and boundaries. The second one encapsulates an important and
often overlooked aspect of communities, namely their locality. A useful example
is to consider an individual and his communities; if two people meet in a remote
area of the network, this should not ripple up to him and affect his communities.

The last point is by far the most important in the construction of the cohesion.
The fundamental principle is linked to the commonly accepted notion that a
community is denser on the inside than towards the outside world, with a twist.

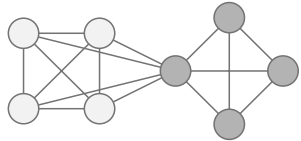
\begin{figure}[htb]
  \centering
  \newcommand{\fourclique}[1]{
    \draw [line width=0.2mm,color=black!55] 
      (0,0) -- (1,0) -- (1,1) -- (0,1) -- (0, 0)
      (1,0) -- (0,1)
      (0,0) -- (1,1) ;
    \foreach \x/\y in {0/0,0/1,1/0,1/1}{
      \draw [line width=0.2mm,color=black!55,fill=#1](\x,\y) circle (0.2) ;
    }
  }
  \begin{tikzpicture}[scale=1]
    \begin{scope}[shift={(-0.25,0)}]
      \draw [line width=0.2mm,color=black!55] 
        (-1.75,-0.5) -- (0.25,0)
        (-1.75, 0.5) -- (0.25,0)
        (-0.75,-0.5) -- (0.25,0)
        (-0.75, 0.5) -- (0.25,0);
    \end{scope}
    \begin{scope}[rotate=-45]
      \fourclique{gray!60}
    \end{scope}
    \begin{scope}[shift={(-2,-0.5)}]
      \fourclique{gray!10}
    \end{scope}
  \end{tikzpicture}
  \caption{Two sets of nodes of identical size, featuring the same number of
  links both inside the set and towards the rest of the network. Despite those
  structural similarities, the darker set appears like a worse community than
  the lighter one.}
  \label{fig:triangles}
\end{figure}

As hinted earlier, the purely vertex/edge based approach to community rating
has flaws. As an example, the toy network in Figure~\ref{fig:triangles}
consists of a group of dark nodes and a group of light nodes. Both groups
contain the same number (4) of nodes and the same number (6) of internal edges
(connecting two nodes in the same group). Moreover, both groups have the same
number (4) of external edges (connecting one node inside the group to one node
outside). That is, with a network vision restricted to nodes and edges, both
groups are virtually indistinguishable, and yet one would say that the dark
group is a ``good'' community, whereas the light group is a ``bad'' community.
The asymmetry between both groups arises when observing triangles -- sets of
three pairwise connected nodes -- in the network: there are 6 outbound
triangles, that is having two vertices inside the dark group and one vertex in
the light group.

The use of triangles does not only stem from the asymmetry they cause in the
treatment of different group but is in line with the notions of \emph{triadic
closure} and \emph{weak ties} introduced by Anatol Rapoport and Mark
Granovetter \cite{Rapoport57,Granovetter:1973wj}. Granovetter defines weak
ties as edges connecting acquaintances, and argues that ``\emph{[\dots] social
systems lacking in weak ties will be fragmented and incoherent. New ideas will
spread slowly, scientific endeavors will be handicapped, and subgroups
separated by race, ethnicity, geography, or other characteristics will have
difficulty reaching a modus vivendi.}''. Furthermore, he states that a
``\emph{weak tie [\dots] becomes not merely a trivial acquaintance tie but
rather a crucial bridge between the two densely knit clumps of close
friends}''.

From there triadic closure is the property on triplets $u,v,w$ that if there
exist a strong tie between $u$ and $v$ and between $u$ and $w$ then there is
at least a weak tie between $v$ and $w$. In the context of complex layered
networks where ties can be of different nature -- \emph{blood-related},
\emph{co-workers} -- one can extend this notion by requiring the two strong
ties to be of the same nature. In that case, when one observes a triangle in a
network, there are chances than the three edges are of the same type, whereas
edges which do not belong to triangles may be considered as weak ties, and as
such serve as a bridge between communities and thus their exclusion from a
community should not be detrimental to its quality. For the same reasons, only
outbound triangles should negatively affect the quality of a group of nodes.

Building on this observation, we now formally define the cohesion. Let
$G=(V,E)$ be a network and $S \in V$ a set of nodes. We define a triangle as
being a triplet of nodes $(u,v,w)\in V^3$ which are pairwise connected,
\emph{ie.} such that $\left((u,v),(v,w),(u,w)\right) \in E^3$. In respect to
$S$, $\tin(S)$ denotes the number of triangles where all nodes belong to $S$
and $\tout(S)$ is the number of outbound triangles of $S$, that is having
exactly two vertices in $S$.

\begin{equation}
  \cohesion(S) = 
  \underbrace{
    \frac{\tin(S)}{{|S| \choose 3}}
  }_{\triangle\text{ ``density''}}
  \times 
  \underbrace{
    \vphantom{\frac{\tin(S)}{{|S| \choose 3}}}
    \frac{\tin(S)}{\tin(S)+\tout(S)}
  }_{\text{isolation}}
  \label{eq:cohesion}
\end{equation}

\begin{figure}[htb]
  \centering
  \begin{tikzpicture}[scale=1]
    

    \draw [line width=0.2mm,color=black!55] (0,1) -- (0,0) -- (1,0) -- (0,1) -- (1,1) -- (2,1) -- (2,0) -- (1,0);
    \draw [line width=0.2mm,color=black!55] (1,0) -- (1,1) -- (2,0);
    \draw [line width=0.2mm,color=black!55,fill=gray!60] (0,0) circle (0.2);
    \draw [line width=0.2mm,color=black!55,fill=gray!60] (0,1) circle (0.2);
    \draw [line width=0.2mm,color=black!55,fill=gray!60] (1,0) circle (0.2);
    \draw [line width=0.2mm,color=black!55,fill=gray!60] (1,1) circle (0.2);
    \draw [line width=0.2mm,color=black!55,fill=gray!10] (2,0) circle (0.2);
    \draw [line width=0.2mm,color=black!55,fill=gray!10] (2,1) circle (0.2);
    
  \end{tikzpicture}
  \caption{In this example, the set of dark nodes contains 4 nodes, features 2 inbound triangles and only 1 outbound triangles, leading to a cohesion $\cohesion=\frac{1}{3}$.}
  \label{fig:cohesion_example}
\end{figure}
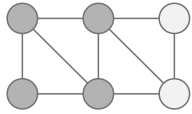

From there, we define the cohesion of $S$ in Equation~\ref{eq:cohesion} as a
product of two factors. The first one is a triangular analog to the usual
definition of density: it denotes the fraction of all possible triangles in a
set of given size $|S|$ which are present in $S$. The second factor is an
isolation factor where, intuitively, a penalty is awarded to the set when there
exist outbound triangles, an example is given on Figure
\ref{fig:cohesion_example}.

The absence of impact of weak ties is naturally encompassed by the definition
of the cohesion: given that the it only relies on counting triangles, deleting
edges which do not belong to any triangles do not affect the number of
triangles and therefore does not impact the value of the cohesion.


\subsection{Evaluation on simple models} 
\label{sub:evaluation_on_simple_models}

\paragraph{Random Networks} 
\label{par:random_networks}
In a random network $G(n,p)$, the expected number of triangles in a set $S_k$
of size $k$ is $\tin = p^3{k \choose 3}$ and the expected number of outbound
triangle is given by $\tout = p^3(n-k){k\choose 2}$. From there, the expected value
(for large $k$ and $n$) of the cohesion is given by $\cohesion(S_k)\sim
p^3\frac{k}{n}$. This exhibits the absence of expected community structure in
random networks as the best possible community is the whole network.

\paragraph{Four groups} 
\label{par:four_groups}
The ``four groups'' test was introduced by Newman and Girvan to test the
accuracy of a community detection algorithm. We here use the same framework to
illustrate the pertinence of the cohesion. The setup is the following:
consider a network of size $4n$ consisting of 4 groups of size $n$. Edges are
placed independently between vertex pairs with probability $p_\text{in}$ for
an edge to fall inside a community and $p_\text{out}$ for an edge to fall
between communities. The cohesion of such a group is given, for large $n$, by
$\cohesion \sim \frac{p_\text{in}^5}{p_\text{in}^2 + 9 p_\text{out}^2}$, which
increases when $p_\text{in}$ increases or $p_\text{out}$ decreases as one
would expect from a quality function.


\section{Fellows} 
\label{sec:Fellows}

Defining a new metric of such a subjective notion as ``how community-like is
this set of nodes ?'' raises the critical issue of its evaluation -- or put
another way, how does one defines the \emph{quality of a quality function}.
While in the previous section we exhibited that the cohesion makes sense on
simple models, this is not enough to validate its use on real data. We now
present Fellows \cite{Fellows:2011}, a large scale online experiment on
Facebook which was conducted in order to provide an empirical evaluation of
the cohesion. The gist of the idea behind Fellows is to quantify the accuracy
of the cohesion by comparing it to subjective ratings given to communities by
real persons.

\subsection{The Experiment} 
\label{sub:The Experiment}

`Fellows' is a single page web application which provides the user with a
short description\footnote{In English, French, Portuguese and Spanish.} of the
experiment and its motivations. When a visitor wishes to take part in the
experiment, they authorize the application to access their personal data on
Facebook. From that point, the application connects to Facebook through the
Facebook API \cite{Facebook:2011} and downloads the list of their friends and
interconnections between pairs of friends to reconstruct the social
neighborhood of the user $\neighbors(u)$. The application also publishes a
message on the user's Facebook wall to invite their friends to participate.
Using a simple greedy algorithm \cite{Friggeri:2011us}, similar in spirit
rather than in metric to one previously introduced by Clauset
\cite{Clauset:2005fm}, the application computes the user's groups of friends
in their immediate social neighborhood by locally maximizing the groups'
cohesion. It is important to note that all computation is done in JavaScript
inside the user's browser and that no \textbf{identifiable} information is
ever transmitted back to the application's server. Statistics on each of the
groups are then sent to the server along with an anonymous unique user and
session identifier (to be able to exclude users participating several times).
The user's and their friends' birthdays and genders are also anonymously
recorded.

\begin{figure}
\centering
\includegraphics[width=\linewidth]{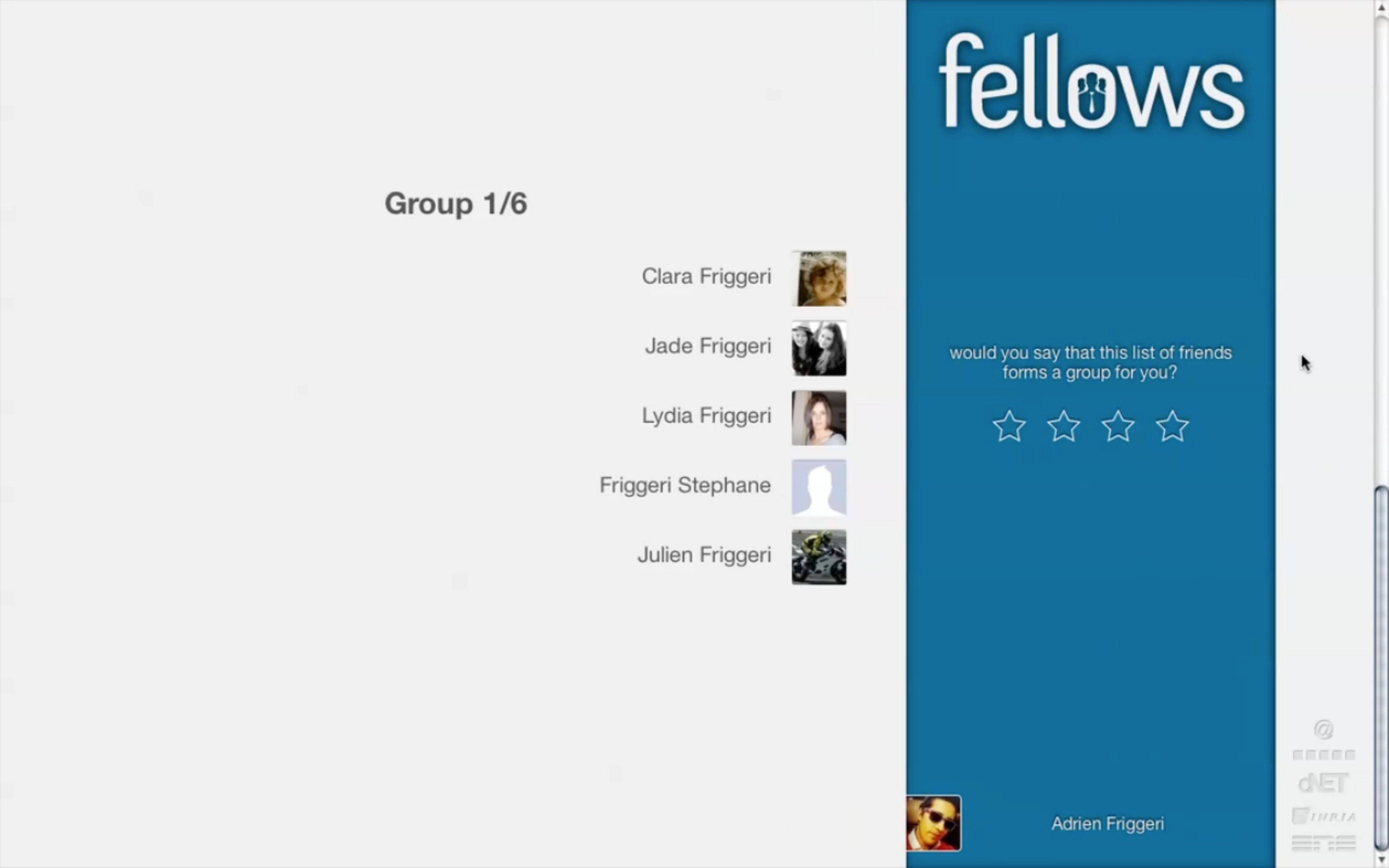}
\caption{Screenshot of the application displaying a group.}
\label{fig:fellows_capture}
\end{figure}

Once those groups are computed, the application displays a list of names and
pictures of friends which are present in the group featuring the highest
cohesion (Fig.~\ref{fig:fellows_capture}). The user is asked to give a
numerical rating between 1 and 4 stars, answering the question ``would you say
that this list of friends forms a group for you?'' They then have the
opportunity to create a Friend List on Facebook, which is a feature which
allows a better control on the diffusion of the information they publish on
the social network. Once they submit the rating, it is uploaded to the
application server where it is associated to the relevant group. In case the
user has created a Friend List, the name they have given is also recorded. The
user is then presented with another group and the process is repeated until
either {\sc i}) the user exits the application or {\sc ii}) all groups are
rated and a message is displayed to thank the user for their involvement.


\subsection{Progress} 
\label{sub:Progress}

\begin{figure}
\centering
\includegraphics[width=\linewidth]{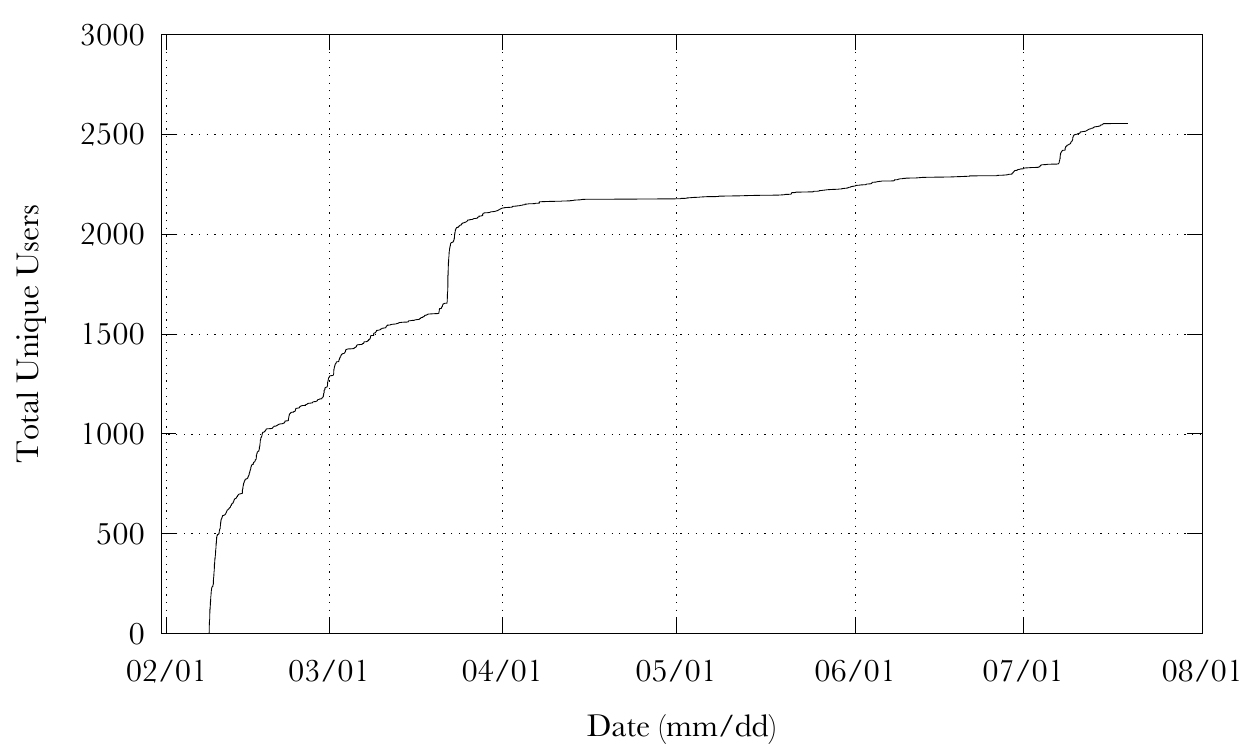}
\caption{Evolution of total unique users through time.}
\label{fig:evolution}
\end{figure}

Fellows was launched on February 8$^\text{th}$, 2011. The authors published a
link to the application on their Facebook walls and sent the {\sc url} to
several active mailing lists. In less than a day, 500 users had taken part in
the experiment and at the time of writing, participations totaled
\FEnbTotalUsers persons (Fig.~\ref{fig:evolution}). Although unrelated to the
evaluation of the cohesion, their are several facts which are interesting in
the spread of the experiment. We observed a pattern of daily increase and
nightly stagnation in the number of participants, corresponding to Western
Europe timezone, which is coherent with data obtained from Google
Analytics\footnote{A service from Google which provides detailed statistics of
visitors access} indicating that the vast majority of Fellows' visitor came
from France.

Moreover, the total number of unique users increases by bursts: observe how on
March 23$^\text{rd}$ the number of users rises from $\sim1700$ to $\sim2000$ in
a single day after having increased by $200$ in two weeks. We have been able to
trace back this sudden influx of participants to the publication of an article
on a high traffic French blog on that date. Although this event was the most
notable, we have been able to manually track down the origin of several
different bursts -- \emph{e.g.} an email on a large mailing list on February
14$^\text{th}$, a tweet by an \emph{influent} twitterer on February
28$^\text{th}$.

As stated above, when a user started the application for the first time, a
message was automatically published on their Facebook wall to invite their
friends to participate. Despite that fact, less than half the incoming traffic
on the website came from Facebook. We conclude unfortunately that either the
message was not appealing enough or that Fellows did not have the same viral
potential as, for example, a double rainbow.


\subsection{Population} 
\label{sub:Population}

\begin{figure}[t]
\centering
\includegraphics[width=\linewidth]{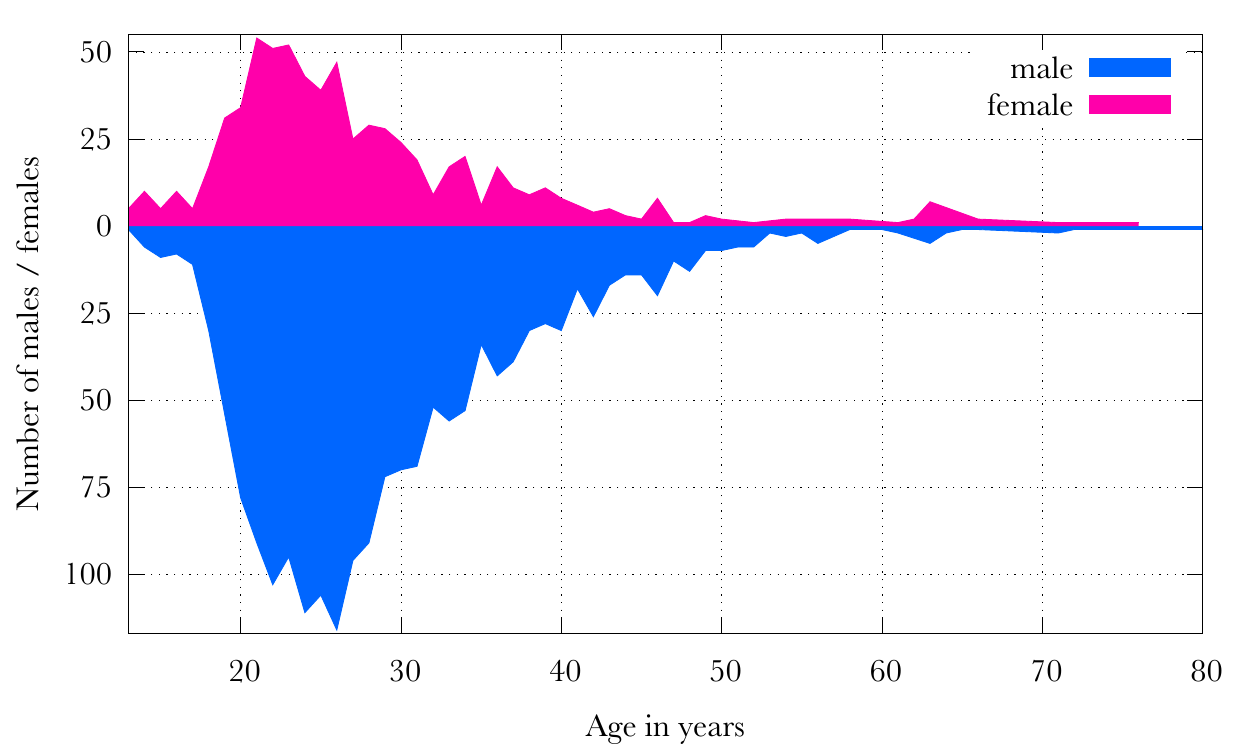}
\caption{Densities of ages of male and female participants.}
\label{fig:gender_ages} 
\end{figure}

In some cases, the participations were corrupted or incomplete -- \emph{e.g.}
the user temporarily lost their internet connection. As a consequence,
\FEnbFailedUsers participations had to be discarded, leaving \FEnbUsers valid
contributions (\FEnbMaleUsers males, \FEnbFemaleUsers females and
\FEnbUnknownUsers persons of unknown gender). The participants were on average
$\FEavAge\pm\FEstdAge$ years old -- male subjects: $\FEavAgeM\pm\FEstdAgeM$
yo, female subjects: $\FEavAgeF\pm\FEstdAgeF$ (age distributions for male and
female subjects are given in Figure~\ref{fig:gender_ages}).

\begin{figure}[b]
\centering
\includegraphics[width=\linewidth]{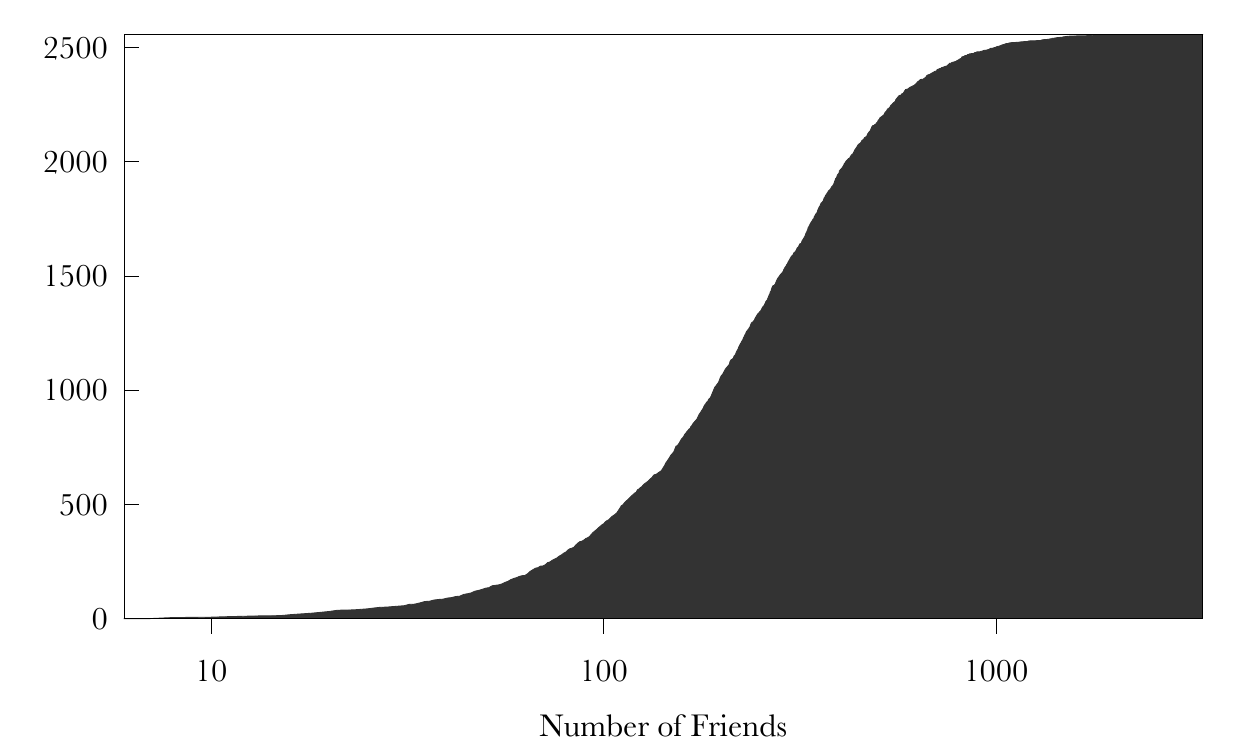}
\caption{Distribution of users' number of friends.}
\label{fig:nbfriends_distrib}
\end{figure}

On Facebook, the number of friends one might have cannot exceed 5000. The
distribution of the number of friends is heterogeneous
(Fig.~\ref{fig:nbfriends_distrib}), with 10\% users having less than
\FEnbFriendsTenPC friends and 90\% users having less than
\FEnbFriendsNinetyPC, the median being at \FEnbFriendsFiftyPC friends.

\section{Experimental Validation} 
\label{sec:Experimental Validation}

\begin{figure}[t]
\centering
\includegraphics[width=\linewidth]{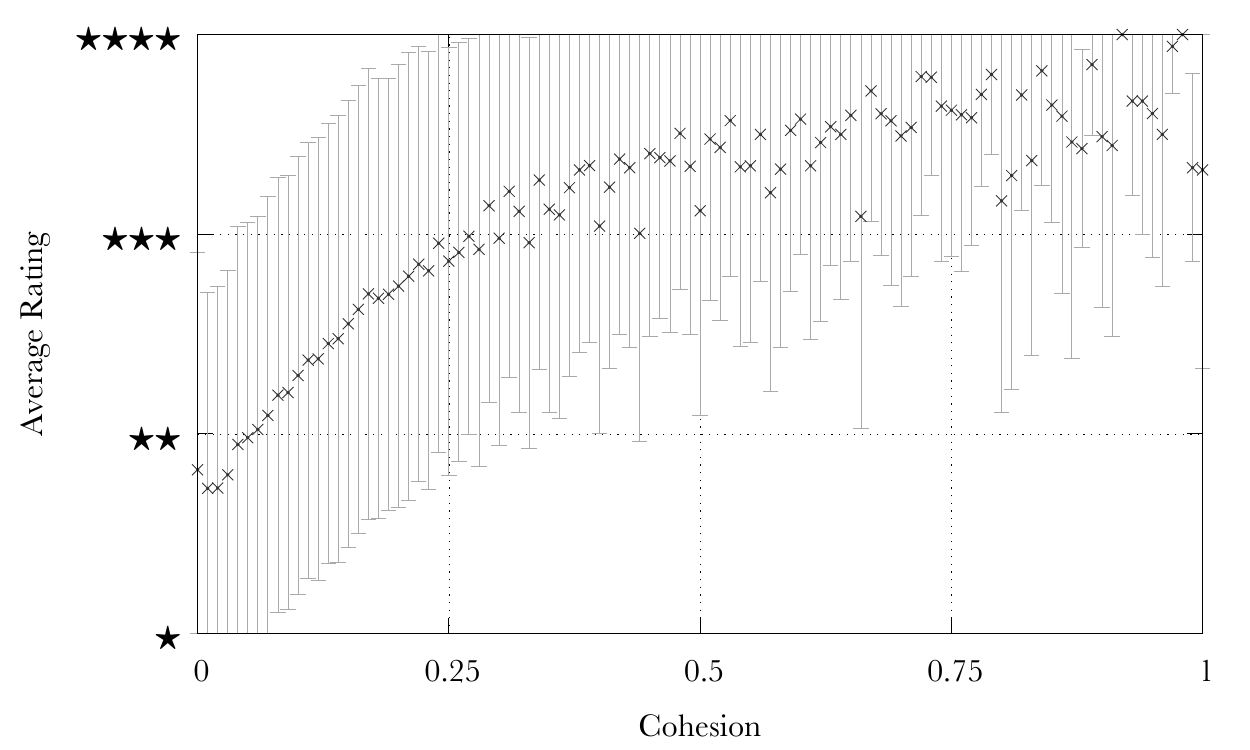}
\caption{Average rating obtained by groups as a function of their cohesion.}
\label{fig:cohesion_rating}
\end{figure}

In this section, we present the main contribution of this article, namely that
the cohesion captures well the community-ness of a set of nodes. We first
present statistics on the ratings which were obtained through the experiment and
then exhibit how both cohesion and ratings are correlated.

\subsection{Ratings overview} 
\label{sub:stars}

The 2157 valid subjects lead to the detection of \FEnbEgomunities groups.
Given the fact that a user could stop the experiment at any time, \FEnbRated
groups received a rating -- however, \FEpercRatedAtLeastNinetyPC of the
subjects rated more than 90\% of their groups. There are several explanations
to those forfeitures, among others: {\sc i}) that the user felt the groups
they were presented with were of poor quality (the non-rated groups have on
average a cohesion $\cohesion = \FEunratedCohesionAv\pm\FEunratedCohesionStd$)
or {\sc ii}) that the user had too many groups to rate -- although the number
of groups is bounded, if a user has a lot of friends, that bound can be
sufficiently high to discourage them.

Out of the 43589 rated groups, \FEpercRatingOne received a rating of 1 star,
\FEpercRatingTwo received 2 stars, \FEpercRatingThree were rated 3 stars and
\FEpercRatingFour were awarded 4 stars. It is important to note here that the
aim of the experiment was \textbf{not} to obtain the highest possible
proportion of 4 stars ratings.

The first thing to notice is that the algorithm assigns all nodes of degree
greater than 3 to at least one group. In practice, there is no reason that all
nodes belong to at least one socially cohesive group: a social neighborhood
might be constituted of an heterogeneous set of communities linked through
weak ties and/or sparse meshes. Moreover, the social topology on Facebook and
in the real world are not isomorphic, not only because people tend to add more
distant acquaintances as Facebook friends, but also due to the presence of
\emph{non-human} profiles representing brands -- incidentally, those would be
better represented as Facebook pages, but for some reasons some organizations
prefer this structure.

Second, and perhaps more important, is that the aim of the experiment is not
to evaluate the quality of the -- rather simple -- algorithm, but that of the
underlying metric. In this context, obtaining low ratings is perfectly
acceptable -- and desirable -- as long as they correlate to the cohesion.

\subsection{Cohesion $\sim \bigstar$} 
\label{sub:cohesionstar}

We now exhibit the experimental links between a structural metric, the
cohesion $\cohesion$, and the subjective appreciation of a group's pertinence
expressed as the average rating $R$ given by users. On
Figure~\ref{fig:cohesion_rating}, we discretize the cohesion of all groups in
increments of 0.01 and we represent the average rating obtained by groups in
the same increment. Both quantities are rank correlated (Spearman's
correlation $\rho = \FEcorCohRatSpearman$, $p\text{-value}=
\FEcorCohRatSpearmanP$). Thus, when the cohesion increases, so does the
average rating, and conversely. Furthermore, $\ln\cohesion$ and $\ln R$ are
linearly correlated (Pearson's correlation $r=\FEcorlogCohlogRatPearson$,
$p\text{-value}=\FEcorlogCohlogRatPearsonP$).

\begin{figure}[t]
\centering
\includegraphics[width=\linewidth]{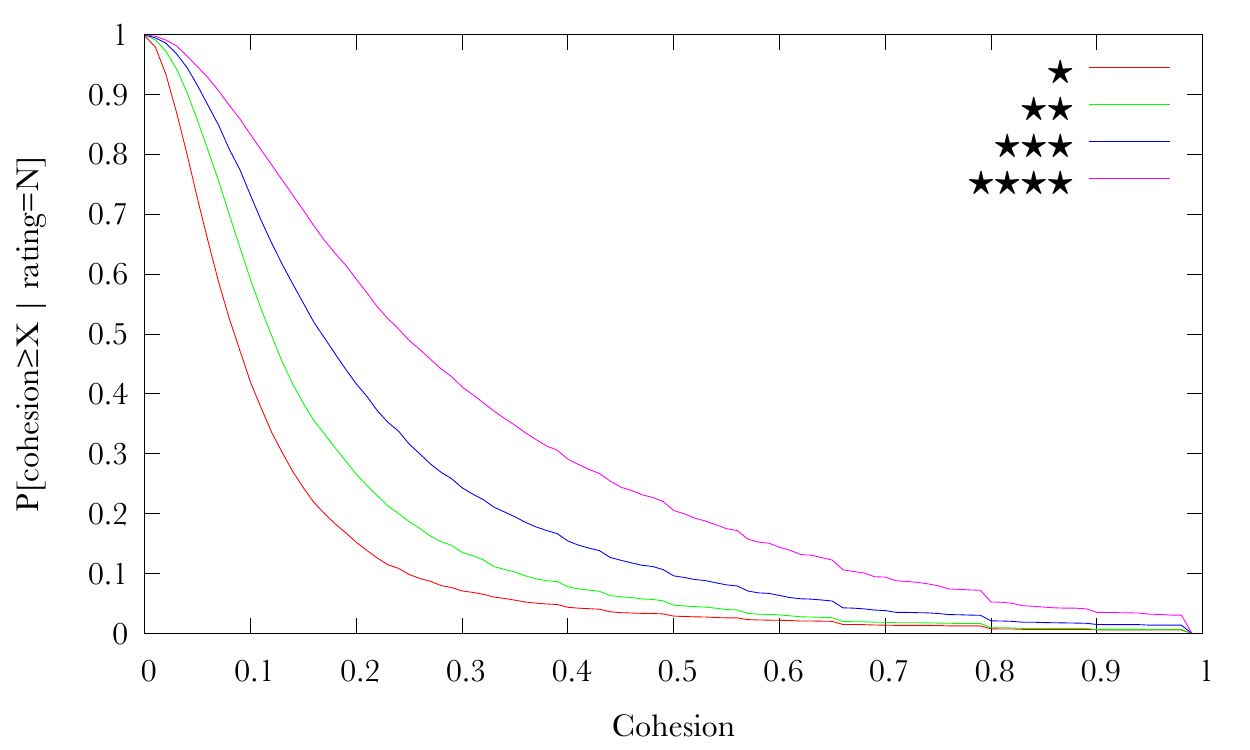}
\caption{Normalized reversed cumulative distribution of cohesion for groups
rated 1,2,3 or 4 stars ($\mathbb{P}[\text{cohesion}\geq X | \text{rating} =
N]).$}
\label{fig:cohesion_distribs}
\end{figure}

On Figure~\ref{fig:cohesion_distribs} we plot the distributions of cohesions
of each of the four sets of groups of rating 1, 2, 3 and 4 stars. From this,
we observe that the higher the rating, the higher the probability of obtaining
high cohesions. Therefore, we conclude that the cohesion in a pertinent
measure to evaluate the community-ness of a set of nodes, as it is highly
correlated to its subjective evaluation.

\begin{figure}[b]
\centering
\includegraphics[width=\linewidth]{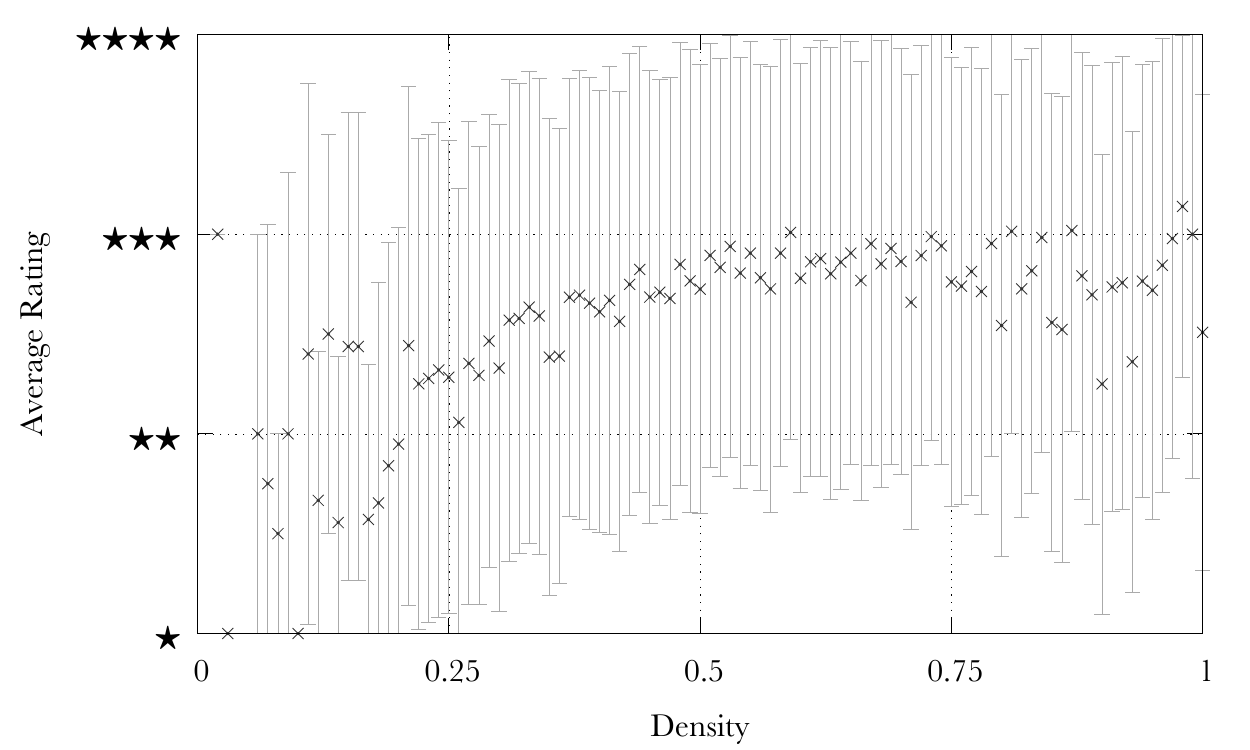}
\caption{Average rating obtained by groups as a function of their density.}
\label{fig:density_rating}
\end{figure}

\begin{figure}[t]
\centering
\includegraphics[width=\linewidth]{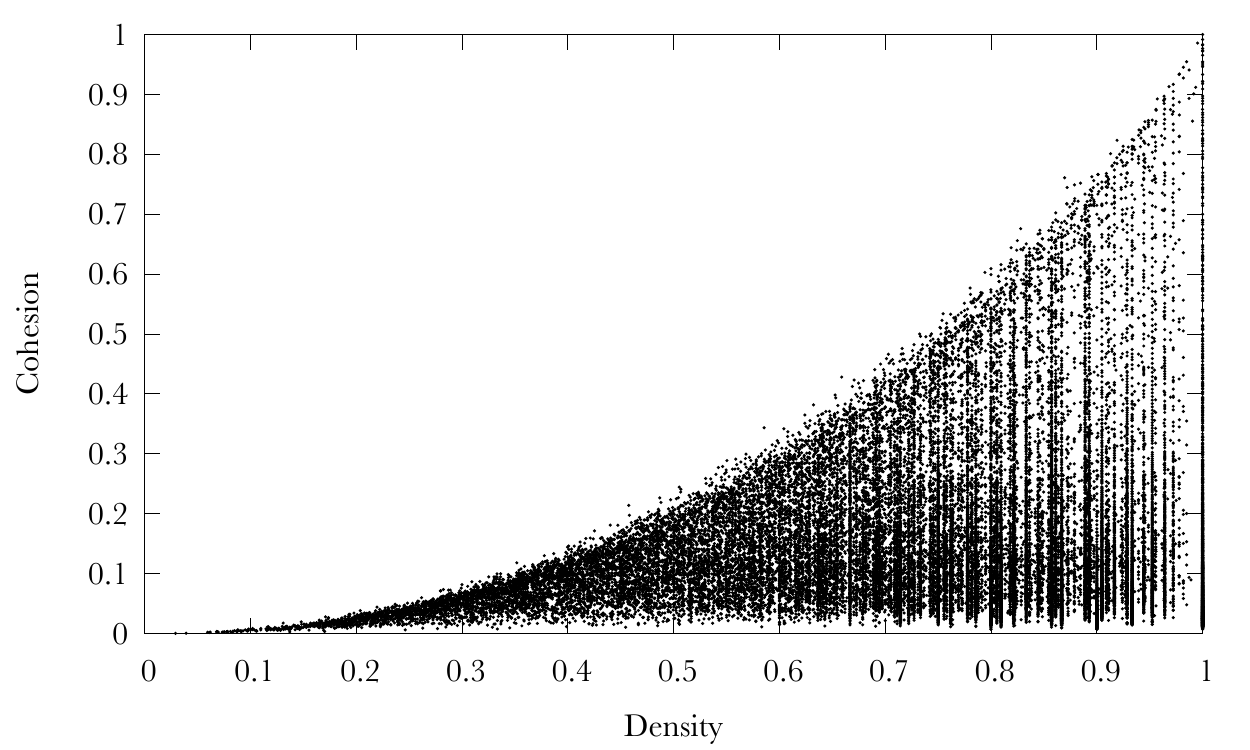}
\caption{Density vs. cohesion.}
\label{fig:density_cohesion}
\end{figure}

Furthermore, it is interesting to look at the relation, if there is any,
between the ratings and other graph metrics, such as the density of the
considered set. On Figure~\ref{fig:density_rating} we plot the average rating
obtained for groups of a given density. Groups having a density greater than
$\frac{1}{3}$ tend to have the same average rating (between 2 and 3 stars).
There seems however that for densities smaller than $\frac{1}{3}$ the rating
increases with the density. To explain this fact, consider that $\cohesion(S)
< \frac{\tin(S)}{{|S| \choose 3}}$. Given that $\tin(S)<m\sqrt{m}$ where $m$
is the number of edges in $S$, there exist a bounding relation between density
and cohesion as exhibited in Figure~\ref{fig:density_cohesion}. Therefore, the
lower ratings obtained by less dense groups can be explained by the fact that
those have low cohesion, which itself is highly correlated to ratings.

For similar reasons, groups having a low clustering coefficient or low
conductance display low ratings, because the clustering coefficient imposes a
higher bound on the number of triangles in the set of nodes and the
conductance imposes a higher bound on the number of outbound triangle. Yet
again, high values of clustering or conductance do not yield high ratings,
because the value of the cohesion can span a far greater range (\emph{e.g.} a
set with high clustering but a lot of outbound triangles might lead to a lower
cohesion than that of a set with lower clustering but lower number of outbound
triangles). As such, we assess that the cohesion leads to a more refined way
of rating communities than by solely considering density, clustering or
conductance.


\section{Ongoing Work} 
\label{sec:ongoing_works}
\subsection{Complexity} 
\label{sub:complexity}

We conjecture that finding a subgraph of maximal cohesion in a given network
is an NP-hard problem and are currently working on a proof. We define the
problem {\sc Subgraph With Cohesion $c$} as follows: Given a graph $G=(V,E)$
and a positive integer $k$, is there a subset $S$ of $V$ such that
$\cohesion(S)=c$. Counter-intuitively, the difficulty seems to arise from low
rather than high cohesion values: here we show that {\sc Subgraph With Cohesion
0} is NP-complete but that {\sc Subgraph With Cohesion 1} can be solved in
polynomial time. The problem for values of $c\in]0,1[$ remains however open.

{\sc Subgraph With Cohesion 0:}
First note that $\cohesion(S)=0$ is equivalent to $\tin(S)=0$, thus the
problem is equivalent to that of finding a triangle-free induced subgraph of
$G$ of size $k$. ``Triangle-free'' is a non-trivial and
hereditary property and as such, per Lewis and Yannakakis \cite{Lewis1980219},
the problem is NP-complete.

{\sc Subgraph With Cohesion 1:}
In this case, a set $S$ such that $\cohesion(S)=1$ is a clique which has 0
outbound triangles. We introduce the notion of triangle connectivity, an
equivalence relation on edges of the network defined as such: two edges $e$
and $e^\prime$ are said to be triangle connected if there exist a sequence of
triangles $(t_i)_{0\leq i \leq N}$ of $G$ such that $e$ is an edge of $t_0$,
$e^\prime$ is an edge of $t_N$ and $\forall i<N, t_i$ and $t_{i+1}$ share a
common edge. From there, if there exist a set $S$ of cohesion 1, then all the
edges of its induced subgraph must be in the same equivalence class and
moreover the equivalence class cannot contain any other edges -- if so, the
associated subgraph would contain an outbound triangle for $S$. In conclusion,
a set of size $k$ with cohesion 1 exists if and only if there is an
equivalence class containing ${k \choose 2}$ edges. Given that it is possible
to list all triangles in polynomial time and that by using a union-find
algorithm one can compute all triangle connected equivalence classes in a time
polynomial in the number of triangles, the problem of finding a set of nodes
of size $k$ having a cohesion 1 can be solved in polynomial time.

\subsection{Extension to weighted networks} 
\label{sub:extension_to_weighted_networks}
Besides complexity analysis, future works will also focus on the evaluation of
weighted cohesion to quantify the quality of weighted social communities. In a
simple unweighted model of social networks, when two people know each other,
their is a link between them. In real life however, things are more subtle, as
the relationships are not quite as binary: two close friends have a stronger bond
than two acquaintances. In this case, weighted networks are a better model to
describe social connections, this is why we deem necessary to introduce an
extension of the cohesion to those networks.

The definition of the cohesion can, as a matter of fact, be extended to take the
weights on edges into account. We make the assumption on the underlying network
that all weights on edges are normalized between 0 and 1. A weight $W(u,v)=0$
meaning that there is no edge (or a null edge) between $u$ and $v$, and a weight
of 1 indicating a strong tie. We define the weight of a triplet of nodes as the
product of its edges weights $W(u,v,w) = W(u,v)W(u,w)W(v,w)$. It then comes 
that a triplet has a strictly positive weight if and only if it is a triangle. We
then define inbound and outbound weights of triangles and finally extend the
cohesion.
\begin{eqnarray*}
  \tin^w(S)      &=& \frac{1}{3}\sum_{(u,v,w) \in S^3} W(u,v,w)\\
  \tout^w(S)     &=& \frac{1}{2}\sum_{u\not\in S, (v,w)\in S^2} W(u,v,w)\\
  \cohesion^w(S) &=& \frac{\tin^w(S)}{{|S| \choose 3}}\times\frac{\tin^w(S)}{\tin^w(S)+\tout^w(S)}
\end{eqnarray*}

\section*{Conclusion} 
\label{sec:Conclusion}

We have presented and justified the introduction of a novel measure, the
cohesion, which quantifies the intrinsic community-ness of a set of nodes of a
given network. We have then confronted the measure to real-world perception
during a large-scale experiment on Facebook and found that the cohesion is
highly correlated to the subjective appreciation of communities of Facebook
users. Moreover, we have shown that there were no correlation between other
metrics such as density and ratings. As such, we conclude that the use of the
cohesion allows a good quantification of the community-ness of a set of nodes.
Future works lie among others in the study of the cohesion from an algorithmic
point of view and extensions to the metric to weighted networks.




\bibliographystyle{IEEEtran}
\bibliography{./biblio}
\end{document}